\documentclass[sigconf, nonacm]{acmart} 
\title[Method of moments for Gaussian mixtures]{Method of moments for Gaussian mixtures: Implementation and benchmarks}

\author{Haley Colgate Kottler}
\orcid{0000-0001-9554-3662}
\affiliation{
    \institution{University of Wisconsin-Madison}
    \department{Department of Mathematics}
    \city{Madison}
    \state{Wisconsin}
    \country{USA}}
\email{haley.kottler@wisc.edu}

 \author{Julia Lindberg}
 \orcid{0000-0002-6840-6053}
 \affiliation{
    \institution{University of Texas at Austin}
    \department{Department of Mathematics}
    \city{Austin}
    \state{Texas}
    \country{USA}}
\email{julia.lindberg@math.utexas.edu}

\author{Jose Israel Rodriguez}
\orcid{0000-0003-3140-9944}
\affiliation{
    \institution{University of Wisconsin-Madison}
    \department{Department of Mathematics}
    \city{Madison}
    \state{Wisconsin}
    \country{USA}}
\email {jose@math.wisc.edu}

\usepackage{latexsym,wasysym, mathrsfs, mathtools,hhline,color, bm}
\usepackage[all, cmtip]{xy}

\usepackage{url}

\definecolor{hot}{RGB}{65,105,225}

\usepackage{ textcomp }

\usepackage{graphicx,enumerate}

\usepackage{enumitem}
\usepackage[normalem]{ulem}

 \usepackage{framed} 

 \usepackage{amsthm}  
 \usepackage{amsmath} 
 \usepackage{amsfonts}  
 \usepackage{amscd} 
 \usepackage[all]{xy} 
\usepackage{graphicx}
\usepackage{pdfsync}
\usepackage[linesnumbered,lined,boxed,commentsnumbered]{algorithm2e}
\RestyleAlgo{ruled}
\usepackage{listings}
\usepackage{cleveref}
\newcounter{theorem}
\newtheorem{thm}{Theorem}[section]

\newtheorem{problem}[thm]{Problem}

\theoremstyle{definition}
\newtheorem{ex}[thm]{Example}

\theoremstyle{definition}

\newtheorem{rem}[thm]{Remark}
\numberwithin{equation}{section}

\newcommand{\R}{\mathbb{R}}
\newcommand{\NN}{\mathbb{N}_{\geq 0}}

\makeatletter
\let\c@table\c@figure 
\let\ftype@table\ftype@figure 
\makeatother

\begin{document}

\begin{abstract}
Gaussian mixture models are universal approximators in the sense that any smooth density can be approximated arbitrarily well with a Gaussian mixture model with enough components.  
Due to their broad expressive power, Gaussian mixture models appear in many applications.
As a result, algebraic parameter recovery for Gaussian mixture models from data is a valuable contribution to multiple fields. 
Our work documents performance of the method of moments for high dimensional Gaussian mixtures. 
We outline the method of moments, and selections of moments and their corresponding polynomials that work well for parameter recovery in practice.  
Our main contribution puts these ideas into practice with an 
implementation as a \texttt{julia} package, \texttt{GMMParameterEstimation}, 
as well as computational benchmarks. 
\end{abstract}

\maketitle

\section{Introduction}
In 1892,  W. F. R. Weldon and Florence Joy Weldon shared data
with Karl Pearson documenting the frequencies of ratios of forehead to body-length among $1000$ crabs. Since the data was skewed, Pearson conjectured that the crab data he was observing came from two subpopulations of crabs and went on to fit 
a convex combination of two Gaussian densities to the data
\cite{pearsonCrabs}. This fitting procedure that Pearson developed is what we now know as the \emph{method of moments}.

Pearson's solution provides one method to address the more general problem of density estimation. Density estimation asks: ``Given a set of samples $\{y_i\}_{i=1}^n$ from an unknown distribution $p$, can we recover $p$?"  For the purpose of this work, we will make the assumption that the underlying distribution is a Gaussian $k$-mixture model.  A random variable $X\in\R^d$ is a \emph{Gaussian} random variable 
if it has density 
$$f_X(x_1,\dots,x_d|\mu,\Sigma) = \frac{1}{\sqrt{(2\pi)^d|\Sigma|}}\exp\left(-\frac{1}{2}(x-\mu)^T\Sigma^{-1}(x-\mu)\right).$$ 
where $\mu\in\mathbb{R}^d$ and $\Sigma\in\R^{d\times d}$ is a symmetric positive definite matrix. We denote $X\sim~N(\mu,\Sigma)$. 
A 
random variable $X \in \R^d$ is distributed as the \emph{mixture of $k$ Gaussians},
$X\sim\sum_{\ell=1}^k\lambda_\ell N(\mu_\ell,\Sigma_\ell)$, if it has density
$$f_X(x_1,\dots,x_d|\lambda_\ell,\mu_\ell,\Sigma_\ell)_{\ell=1,\dots,k} = \sum_{\ell=1}^k\lambda_\ell f_{X_\ell}(x_1,\dots,x_n|\mu_\ell,\Sigma_\ell)$$
where $(\lambda_1,\dots,\lambda_d)\in\Delta_{k-1}=\{\lambda\in\R_{>0}^k:\sum_{\ell=1}^k\lambda_\ell=1\}$, $\mu_\ell\in\R^d$, and $\Sigma_\ell\in\R^{d\times d}$ is symmetric and positive definite for $\ell\in[k]$.  

Gaussian mixture models are a popular class of densities due to their broad expressive power. Specifically, Gaussian mixture models are \emph{universal approximators} in the sense that any smooth density can be approximated arbitrarily well with a Gaussian mixture model with enough components \cite[Chapter 3]{Goodfellow-et-al-2016}. Due to their broad expressive power, Gaussian mixture models appear in many applications \cite{1402510, 10.1007/s00158-020-02676-3, 4656564}.  

There has been much work on density estimation for Gaussian mixture models. A popular approach is \emph{maximum likelihood estimation}. Maximum likelihood estimation is a method of density estimation that seeks to maximize the likelihood that a given set of samples comes from a certain density. In the context of Gaussian mixture models, this problem is stated as:
\begin{align}
   \text{argmax}_{\mu, \sigma^2, \lambda} \ \  \prod_{j=1}^n  \sum_{\ell=1}^k \lambda_\ell \frac{1}{\sqrt{2 \pi \sigma_\ell^2}} \exp \Big( - \frac{(y_j - \mu_\ell)^2}{2\sigma_\ell^2} \Big).\label{mle}
\end{align} 

Whenever $k \geq 2$, 
\eqref{mle} is nonconvex so finding a global solution is difficult. A common algorithm used in practice is called expectation maximization (EM). The EM algorithm finds local maxima of \eqref{mle} by treating Gaussian mixture models as latent variable models and successively updating the weights $\{\lambda_\ell\}_{\ell=1}^k$ then the mixture parameters $\{\mu_\ell, \Sigma_\ell \}_{\ell=1}^k$  \cite{dempster1977maximum}. 

There are numerous downsides of the EM algorithm including that it is sensitive to the starting point used to initialize the algorithm and that it requires all samples at each iteration, which can be prohibitive for large data sets. Despite these downsides, there has been much recent work analyzing the performance of the EM algorithm in a variety of regimes. For instance, when $\lambda_1 = \lambda_2 = \frac{1}{2}$ and all of the covariance matrices are known and equal, it has been shown that the EM algorithm converges geometrically to the global optimum \cite{daskalakis2017ten,xu2016global}.
Other work has studied the global landscape of the EM algorithm and the structure of local optima in this setting  \cite{chen2020likelihood,xu2016global}. 

We consider a strictly more general problem than density estimation.
In this paper we consider the problem of \emph{parameter recovery}, meaning we wish to find parameters $\{\lambda_\ell, \mu_\ell, \Sigma_\ell \}_{\ell = 1}^k$ close to the true parameters of the underlying Gaussian mixture model. This ensures that the results are more interpretable than those obtained from doing density estimation since
if parameters are recovered, 
the components can be analyzed to learn more about subgroups in the data. Formally, 
we state our problem as follows.
\begin{problem}\label{prob:param_recovery}
    Given samples $\{y_i\}_{i=1}^n$ from a $d-$dimensional Gaussian $k$-mixture model $X\sim\sum_{\ell=1}^k\lambda_\ell N(\mu_\ell,\Sigma_\ell)$, recover $\lambda_\ell,\mu_\ell,\Sigma_\ell$ for $\ell\in[k]$.
\end{problem}

In this paper we propose a solution to \Cref{prob:param_recovery} using the method of moments. As highlighted earlier, this methodology was first proposed and resolved for the mixture of two univariate Gaussians by Karl Pearson in $1892$ \cite{pearsonCrabs}. 
This approach was revisited in 2010 in a series of papers \cite{kalai2010efficiently,moitra2010settling}  
where an algorithm that scales polynomially in $d$ and the number of samples required scales polynomially in the desired accuracy was presented.
{To provide additional background, the method of moments has received attention in algebraic statistics and computational algebra community from several perspectives. 
This includes identifiability for Gaussian mixtures~\cite{ALR2021-decomposable,Amendola-Ranestad-Sturmfels-identifiability-GMM} and inverse Gaussians~\cite{Henriksson-Seccia-Yu-inverse-gaussian}, as well as understanding moment varieties~\cite{Amendola-Faugere-Sturmfels-moment-varieties,Kohn-Sturmfels-2020-polytopes,Amendola-Drton-Grosdos-Homs-Robeva-third-order,Yulia-Kileel-Sturmfels-nonparametric}.
}

We consider both exact and sample moments, since exact moments are believed to appear in cryogenic electron microscopy \cite{sharon2020method}, and sample moments are more commonly used.

\subsection*{Contributions}
\begin{figure*}[h!]
    \centering
    \includegraphics[width=.95\linewidth]{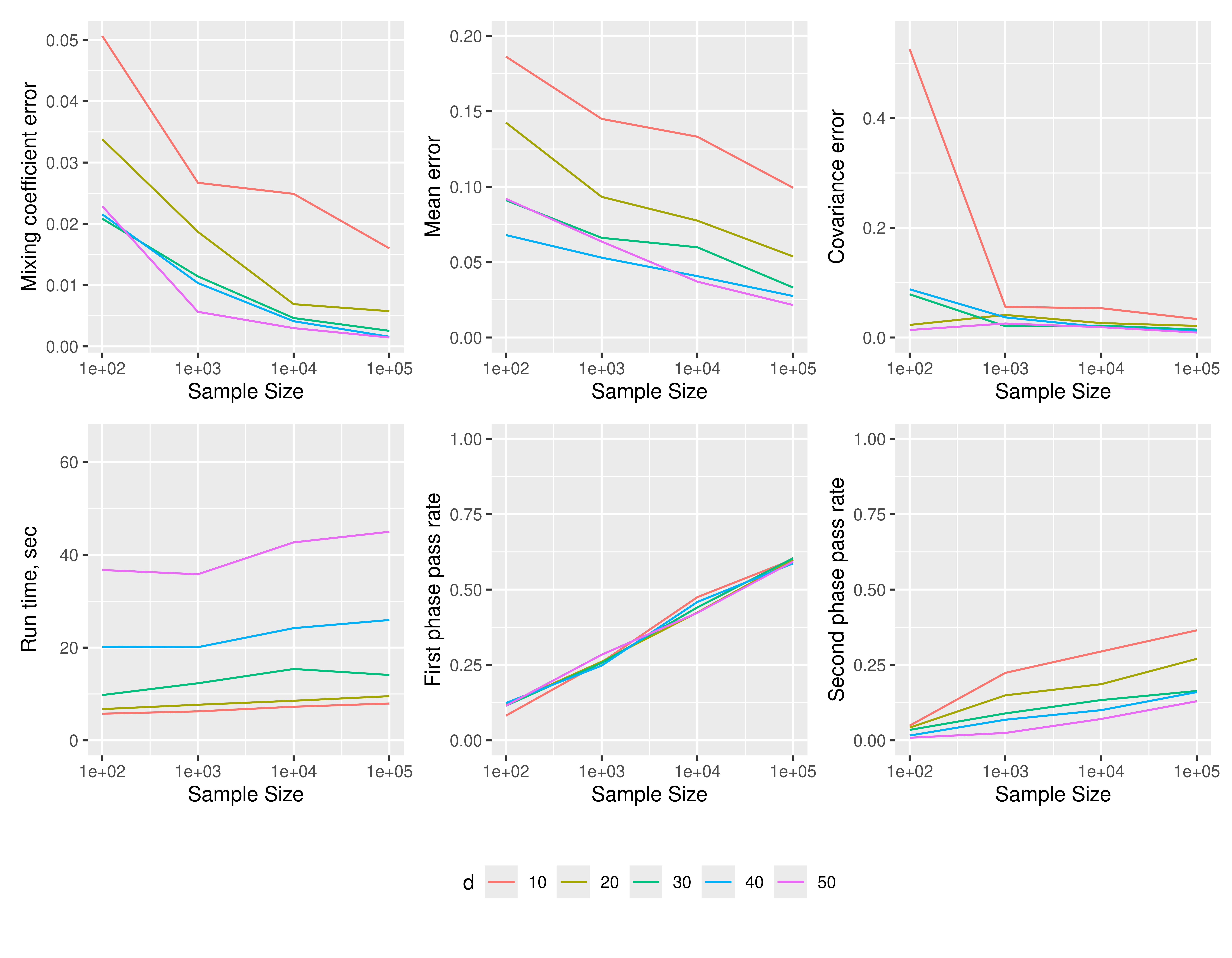}
    \Description{Graphs showing decreasing errors and increasing pass rates and run times as sample size increases}
    \caption{(Sample moments, $\Sigma_\ell\succ0$)
Median mixing coefficient, mean, and covariance errors, and number of statistically meaningful solutions for sample moments over 1000 random test runs with unknown mixing coefficients with $k = 3$ normalized by the number of parameters.}\label{fig:sample-moments-general-cov-unknown-mixing}
\end{figure*}

In this paper we focus on the method of moments approach that provides exact solutions as proposed in \cite{lindberg2023estimating}. This work extends the algorithm implementation to models with non-diagonal covariance matrices. 
Our main contribution is highlighted by the computational results and benchmarks in Section~\ref{sec:computational}:
\begin{itemize}[leftmargin=*]
    \item We provide open source implementations to solve the parameter recovery problem for several classes of Gaussian mixture models.
    \item 
    Our implementation includes several options to compare the use of different choices of moment equations for parameter recovery in the multivariate Gaussian setting (Figure~\ref{fig:kvslowsystems}). 
    \item  We achieve machine precision with exact moments (Fig.~\ref{fig:perfect-moments-general-cov-known-mixing} and ~\ref{fig:perfect-moments-general-cov-unknown-mixing}).
    \item Our implementation is flexible, and includes multiple options to systematically study several classes of Gaussian mixtures; e.g., known and unknown mixing coefficients~(Figures \ref{fig:sample-moments-general-cov-unknown-mixing} and ~\ref{fig:sample-moments-general-cov-known-mixing}).
\end{itemize}

\section{Preliminaries and Implementation}
In this section we present some preliminary statistics and algebra background.

\subsection{Univariate Method of Moments}
For a random variable $X \in \mathbb{R}$ with density $f: \mathbb{R} \to \mathbb{R}$ and any $i \geq 0$, the $i$-th moment of $X$ is defined as $$m_i=E[X^i]=\int x^i f(x) dx.$$  

\begin{ex}
For $X\sim\mathcal{N}(\mu,\sigma)$, the moments are polynomials in $\mu$ and $\sigma$. The first three moments are
$$
    m_0 = 1,\quad\quad
    m_1 = \mu,\quad\quad
    m_2 = \mu + \sigma.
    $$
For $i\geq 2$ the moments satisfy the recursion $m_i=\mu m_{i-1}+(i-1)\sigma m_{i-2}$.
\hfill$\diamond$\end{ex}

For a univariate Gaussian mixture model $X\sim\sum_{\ell=1}^k\lambda_\ell\mathcal{N}(\mu_\ell,\sigma_\ell)$, the moments are given by convex combinations of the Gaussian distribution moments.  For example, 

\begin{align*}
    m_0 &= \sum_{\ell=1}^k\lambda_\ell, &
    m_1 &= \sum_{\ell=1}^k\lambda_\ell\mu_\ell, &    m_2 &= \sum_{\ell=1}^k\lambda_\ell(\mu_\ell^2+\sigma_\ell).
\end{align*}
Moving forward, we will denote the polynomial associated with moment $m_i$ as $f_i(\theta)$ where $\theta$ is the parameter set $\{\mu_\ell,\Sigma_\ell\}_{\ell=1}^k$.

For samples $\{y_1,y_2,\dots,y_N\}$ from the distribution of $X$, the $i$th sample moment is defined as $$\overline{m_i}=\frac{1}{N}\sum_{j=1}^N y_j^i.$$  The sample moments approach the exact moments 
as $N\rightarrow\infty$, so by setting the moment polynomials equal to the empirical moments, we can then solve the polynomial system to recover the parameters.  If the parameters can be identified from the first $M$ moments, then we seek to solve the polynomial system of equations, 
\begin{equation}\label{eq:univariate-moment-system}
f_i(\theta)-\overline{m_i}=0 \text{ for } i\in[M].
\end{equation}

\begin{ex}
    From the set of data on the ratio of forehead to body-length of crabs \cite{pearsonCrabs}, the system is as follows: 

    \begin{align*}
        \lambda_1+\lambda_2 &= 1\\
        \lambda_1\mu_1 + \lambda_2\mu_2 &= 17\\
        \lambda_1(\mu_1^2+\sigma_1) + \lambda_2(\mu_2^2+\sigma_2) &= 305\\
        \lambda_1(\mu_1^3 + 3\mu_1\sigma_1) + \lambda_2(\mu_2^3 + 3\mu_2\sigma_2) &=  5832\\
        \lambda_1(\mu_1^4 + 6\mu_1^2\sigma_1 + 3\sigma_1^2) + \lambda_2(\mu_2^4 + 6\mu_2^2\sigma_1 + 3\sigma_2^2) &= 116061\\
        \lambda_1(\mu_1^5 + 10\mu_1^3\sigma_1 + 15\mu_1\sigma_1^2) + \lambda_2(\mu_2^5 + 10\mu_2^3\sigma_2 + 15\mu_2\sigma_2^2) &= 2385610.
    \end{align*}
\noindent
This can be solved via the following code snippet:
\begin{leftbar}
\begin{verbatim}
using GMMParameterEstimation
d = 1
k = 2
mV1 = [1.0, 16.799, 304.923, 5831.759, ...
estimate_parameters(d, k, mV1, zeros(0,5))
\end{verbatim}
\end{leftbar}

One possible solution is $\lambda_1=0.58$, $\lambda_2=0.42$, $\mu_1=19.30$, $\mu_2=13.40$, $\sigma_1=9.67$, and $\sigma_2=20.35$.  Note, unlike the numbers reported by Pearson, this solution is given based on the abscissae without scaling back to millims.
\hfill$\diamond$\end{ex}

\subsection{Multivariate Method of Moments}
This process is generalized to the multivariate Gaussian mixture setting as follows.
For a multivariate random variable $X\in\R^d$ with density $f_X$ the moments are defined as 
\begin{align*}
m_{v_1,\dots,v_d} &= E[X_1^{v_1}\cdots X_d^{v_d}]\\ &= \int\cdots\int x_1^{v_1}\cdots x_d^{v_d}f_X(x_1,\dots,x_d)dx_1\cdots dx_d
\end{align*}
and the empirical moments are defined as $$\overline{m}_{v_1,\dots,v_d} = \frac{1}{N}\sum_{j=1}^Ny_{j_1}^{v_1}\cdots y_{j_d}^{v_d}.$$  The moment $m_{v_1,\dots,v_d}$ has order $\sum_{j=1}^d v_j$.  For a multivariate Gaussian mixture model,
$m_{v_1,\dots,v_d}$ is a 
polynomial $f_{v}(\theta)$ in the parameters $\theta=\{\lambda_\ell,\mu_\ell,\Sigma_\ell\}_{\ell=1}^k$ where $\mu_j\in\mathbb{R}^d$ and $\Sigma_j\in\mathbb{R}^{d\times d}$ is positive definite.  

When we append
an additional index to the entries of the means and covariances, 
the first index refers to the mixture component.
For example, $\mu_{1j}$ is the $j$-th entry of the mean of component 1.

By setting the polynomials equal to the empirical moments, we formulate this system of polynomials: 
\begin{equation}\label{eq:multivariate-moment-system}
\left\{f_v(\theta)-\overline{m_{v}}=0 \quad \text{ such that }\quad  v\in V\right\},
\end{equation}
where $V$ is a set 
of suitable vectors in $\NN^d$
used to index the moment equations.  

Recovery of the parameters
of a Gaussian mixture model 
using the method of moments depends on the choice of $V$. 
Our implementation views $V$ as the union of three disjoint sets. 
The first, $V_1$, is used to identify the mixing coefficients, a mean and a variance. The second, $V_2$, identifies the other means and variances, while $V_3$ is used to identify the remaining entries of the covariance matrix. 

\begin{rem}{\bf{Notation:}}
To describe the choices in our implementation, 
let $e_i$ denote the standard unit vector with $i$th coordinate $1$ and $d$ entries.
{Note,} for $j\in[d]$, $m_{e_j}$ equals the $j$-th order moment of the univariate Gaussian mixture model given by 
$\sum_{\ell=1}^k\lambda_\ell\mathcal{N}(\mu_{\ell j},\sigma_{\ell j j})$.  
Due to this identification, 
by letting
\[V_1:=\{e_1,2e_1,\dots,(3k)e_1\}\subset\NN^d,
\]
the moment equations 
$\{m_v-\overline{m}_v=0:v\in V_1\}$ can be used to identify $\lambda_\ell,\mu_{\ell 1},$ and $\sigma_{\ell11}$ for $\ell\in[k]$.  
Having
\[V_2:=\{e_i,\dots,(2k+1)e_i:2\leq i\leq d\}\subset\NN^d\]
the set of moments $\{m_v - \overline{m}_v:v\in V_2\}$ 
can then be used to determine 
$\mu_{\ell i}$ and $\sigma_{\ell ii}$ for $\ell\in[k]$ and $2\leq i\leq d$.  Determining the remaining covariance entries will be treated in Section \ref{offdiag}.
\end{rem}
\subsection{Statistically Meaningful Solutions}
Many of the solutions to the systems of polynomials we describe are impractical in a statistical context.  In application, parameters must be real numbers, variances and mixing coefficients must be positive, and covariance matrices must be symmetric and positive definite.  We filter out any solutions that do not meet the requirements for real numbers, positive variances and positive mixing coefficients.  Covariance matrices are symmetric by design.  We then take the eigendecomposition of each covariance matrix, replace any negative eigenvalues with a small normal perturbation, and recompose the matrix to achieve positive definite covariance matrices.

\begin{ex}\label{ex:visualizing-data-and-implementation}

\begin{figure}[hbt]
    \centering
    \includegraphics[width=.8\linewidth]{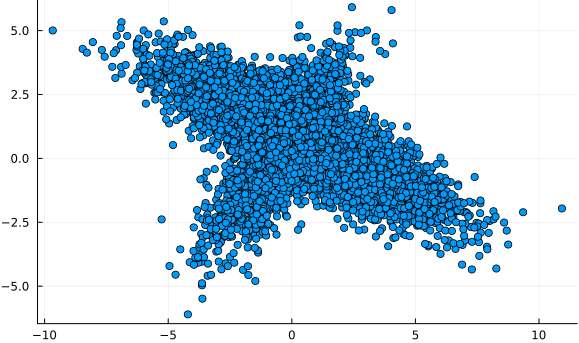}
    \Description{Sample from a 2-dimensional Gaussian 2-mixture model, $n=10000$ forming two overlapping ellipses with similar centers but perpendicular axes}
    \caption{Sample from a 2-dimensional Gaussian 2-mixture model, $n=10000$}
    \label{fig:d-equal-2-data}
\end{figure}

For $d=2$, $k=2$, we solve the moment system
\eqref{eq:multivariate-moment-system} where $V =V_1\cup V_2$ from the data in Figure~\ref{fig:d-equal-2-data}.  This is a sample from $$0.2N\left(\begin{bmatrix} -0.6\\ 0.5\end{bmatrix}, \begin{bmatrix}1.7&1.3\\1.3&3.2\end{bmatrix} \right)+0.8N\left(\begin{bmatrix} 0.3\\ 0.7\end{bmatrix},  \begin{bmatrix}6.1&-2.8\\-2.8&1.9\end{bmatrix} \right).$$
From this data we have the empirical moments 
\begin{align*}
m=[&0.266, 5.444, 6.473, 99.398, 214.853, 3126.467,\\
&0.624, 2.583, 4.055, 19.880, 44.246, 1.130, -1.790].
\end{align*}
Using our implementation we recover a set of parameters for a statistically meaningful solution.

\begin{leftbar}
\begin{verbatim}
using GMMParameterEstimation
d = 2
k = 2
mV1 = [1, 0.266, 5.444, 6.473, 99.398, 214.853, 3126.467]
mV2 = [0.624 2.583 4.055 19.880 44.246]
mV3 = Dict([2, 1] => 1.130, [1, 1] => -1.790)

estimate_parameters(d, k, mV1, mV2, mV3)
\end{verbatim}
\end{leftbar}

This results in $\lambda = [0.21, 0.79]$, $\mu_1=\begin{bmatrix} -0.50\\ 0.47\end{bmatrix}$, $\mu_2=\begin{bmatrix} 0.26\\ 0.72\end{bmatrix}$, $\Sigma_1=\begin{bmatrix}1.55&1.73\\1.73&3.12\end{bmatrix}$, and $\Sigma_2=\begin{bmatrix}6.19&-3.03\\-3.03&1.90\end{bmatrix}$.

\hfill$\diamond$\end{ex}

\section{Parameter Recovery with Non-Diagonal Covariance Matrices}\label{offdiag}

\begin{algorithm}
\SetKwInOut{Input}{Input}\SetKwInOut{Output}{Output}
\caption{Density Estimation for Mixtures of Multivariate Gaussians\cite{lindberg2023estimating}}\label{alg:general}
\Input{The set of sample moments: $V_1,V_2,V_3$ that are the moments to multivariate Gaussian mixture model: $$\lambda_1\mathcal{N}(\mu_1,\Sigma_1)+\cdots+\lambda_k\mathcal{N}(\mu_k,\Sigma_k).$$}
\Output{Parameters $\lambda_\ell\in\mathbb{R}$, $\mu_\ell\in\mathbb{R}^d$, $\Sigma_\ell\succ0$ for $\ell\in[k]$ such that $V_1$, $V_2$, $V_3$ are the moments of distribution $\sum_{\ell=1}^k\lambda_\ell\mathcal{N}(\mu_\ell,\Sigma_\ell)$.}
\BlankLine
Solve the general univariate case using sample moments $V_1\setminus \{\overline{m}_{(3k)e_1}\}$ to get parameters $\lambda_\ell$, $\mu_{\ell,1}$, and $\sigma_{\ell,1,1}$\;
Select the statistically meaningful solution with sample moment $\overline{m}_{3ke_1}$.\;
Solve the univariate case $n-1$ times using the mixing coefficients $\lambda_\ell$ and sample moments $V_2\setminus\{\overline{m}_{(2k+1)e_i}\}_{i=2}^d$ to obtain $\mu_{\ell_i}$ and $\sigma_{\ell ii}$ for $\ell\in[k],2\leq i\leq d$\;
Select the statistically meaningful solution with sample moment $\overline{m}_{(2k+1)e_i}$ for $2\leq i\leq d$\;
Using $V_3$, solve the remaining system of linear equations\;
Return $\lambda_\ell,\mu_\ell,\Sigma_\ell,\ell\in[k]$\;
\end{algorithm}

We implement and numerically test the state of the art algorithm for parameter recovery of Gaussian mixture models presented in \cite{lindberg2023estimating}. The authors of this paper provide detailed information about which moments to use to recover the parameters of a Gaussian mixture model 
with computational experiments to justify its effectiveness.

The effectiveness of selecting the moments for 
the off-diagonal implementation was left to the practitioner. 
Our computational results in the next section give empirical evidence for this selection of moments.  
 
Our implementation provides two options for moment selection. For the the first one, Lindberg et al., demonstrated that for fixed $i,j\in[d]$, $i\neq j$, the set of moment equations $V_3 = \{m_{te_i+e_j}\}_{t=1}^k$ are linear in $\sigma_{\ell i j}$ for $\ell\in[k]$ and leads to a linear system that generically has a unique solution.  Due to the symmetry of covariance matrices ($\sigma_{\ell i j}=\sigma_{\ell j i}$) we add the restriction that $i<j$.  

The second one leads to a system involving lower order moments.
Specifically, by further relying on the symmetry of the covariance matrices and their corresponding polynomials,
we present a lower order system involving these moments
$$V_3 = \begin{cases}
    \{m_{te_i+e_j}, m_{e_i+te_j}\}_{t=1}^{\frac{k}{2}}\cup\{m_{te_i+e_j}\}_{t=1}^{\frac{k}{2}+1} & \text{ if $k$ is even}\\
    \{m_{te_i+e_j}, m_{e_i+te_j}\}_{t=1}^{\frac{k+1}{2}} & \text{ if $k$ is odd}
    \end{cases},$$ where we again assume $i,j\in[d]$ with $i<j$.

\begin{figure*}
    \centering
    \includegraphics[width=.95\linewidth]{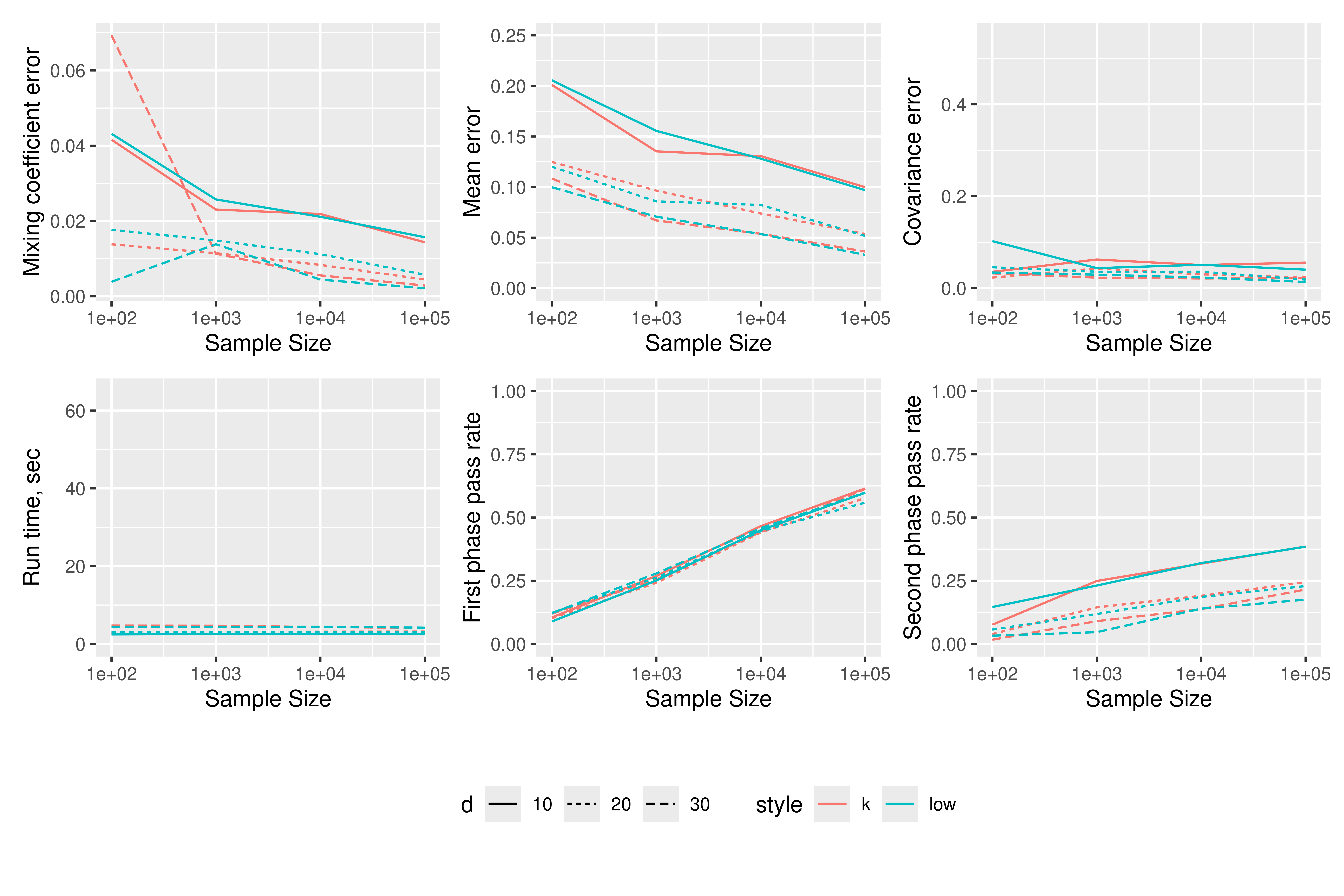}
    \Description{Graphs showing decreasing errors and increasing pass rates with two different systems of equations for the off diagonal entries that trade off which is more effective depending on sample size}
    \caption{(Sample moments, unknown mixing coefficients, comparison of the two recursive systems) Median mixing, mean, and covariance errors, timings, and pass rates over 1000 random test runs with unknown mixing coefficients normalized by the number of parameters.}
    \label{fig:kvslowsystems}
\end{figure*}

As shown in Figure \ref{fig:kvslowsystems}, it is not immediately apparent which system is more efficient or effective for solving this problem.  This distinction may become clearer with more mixture components.

The additional requirement of $i<j$ is preferable because the first dimension is used to determine the mixing coefficients if they are unknown which carry over into the remaining computations.  Accuracy in this dimension is therefore desirable, and if this dimension is in practice more accurate than the others than better estimations will occur when using higher powers here as opposed to other dimensions when possible.  Our updated algorithm is then Algorithm \ref{alg:general}.

Given a $d\times n$ matrix of samples,
the following code
will return the moments necessary for solving the first diameter, the moments necessary for solving the remaining diagonal covariance entries and means, and the moments for the Lindberg et al. system.  
\begin{leftbar}
    \texttt{sampleMoments(sample, k; method = "k")}
\end{leftbar} 
Similarly, 
\begin{leftbar}
\texttt{sampleMoments(sample, k; method = "low")}, 
\end{leftbar}
\noindent will return the moments necessary for solving the first diameter, the moments necessary for solving the remaining diagonal covariance entries and means, and the moments for the lower order system. Our primary function, \texttt{estimate\_parameters}, automatically detects which of these two systems are input for the off-diagonal system and solves accordingly.

Observe that the polynomial $f_{v}(\theta)$ can be linear in the off diagonal entries of $\Sigma_1,\ldots,\Sigma_k$ for other moment equations aside from the aforementioned. For instance, when $|\{j \ : \ v_j = 1 \}| =3$ and the remaining $v_j = 0$, the polynomial $f_{v}(\theta)$ is linear in the off diagonal entries of $\Sigma_1,\ldots,\Sigma_k$. For simplicity of presentation and due to the computational benefit of using lower order moments, we did not consider these systems.

\begin{algorithm}[htb]
\SetKwInOut{Input}{Input}\SetKwInOut{Output}{Output}
\caption{Density Estimation for Mixtures of Multivariate Gaussians}\label{alg:general_cycling}
\Input{The set of sample moments: $V_1$, $V_2$, $V_3$ that are the moments to multivariate Gaussian mixture model: $$\lambda_1\mathcal{N}(\mu_1,\Sigma_1)+\cdots+\lambda_k\mathcal{N}(\mu_k,\Sigma_k)$$.}
\Output{Parameters $\lambda_\ell\in\mathbb{R}$, $\mu_\ell\in\mathbb{R}^d$, $\Sigma_\ell\succ0$ for $\ell\in[k]$ such that $V_1$, $V_2$, $V_3$ are the moments of distribution $\sum_{\ell=1}^k\lambda_\ell\mathcal{N}(\mu_\ell,\Sigma_\ell)$.}
\BlankLine
Set $n=1$\;
Solve the general univariate case using sample moments $V_1\setminus \{\overline{m}_{(3k)e_1}\}$ to get parameters $\lambda_\ell$, $\mu_{\ell,n}$, and $\sigma_{\ell,n,n}$\;
Select the statistically meaningful solution with sample moment $\overline{m}_{3ke_1}$. If none exist, $n = n+1$ and return to Step 1\;
Solve the univariate case $n-1$ times using the mixing coefficients $\lambda_\ell$ and sample moments $V_2\setminus\{\overline{m}_{(2k+1)e_i}\}_{i=2}^d$ to obtain $\mu_{\ell_i}$ and $\sigma_{\ell ii}$ for $\ell\in[k],2\leq i\leq d$ where $i\neq n$\;
Select the statistically meaningful solution with sample moment $\overline{m}_{(2k+1)e_i}$ for $2\leq i\leq d$ where $i\neq n$. If none exist, $n = n+1$ and return to Step 1\;
Using $V_3$, solve the remaining system of linear equations\;
Return $\lambda_\ell,\mu_\ell,\Sigma_\ell,\ell\in[k]$\;
\end{algorithm}
\subsection{Off-Diagonal System Generation}

With the problem of moment selection satisfactorily treated, we move on to generation of the polynomial systems.  
We implemented two methods, one using tensor moments and one using recursion.  This exhibits a contribution of this software in that it  enabled the first replicable comparison of two different methods.

\subsubsection{Tensor Moment Method}

To generate the tensor moments, we use the closed form formula for the mixed dimensional moments of a multivariate Gaussian that is provided by Jo\~{a}o M. Pereira, Joe Kileel, and Tamara G. Kolda in
\cite{pereira2022tensor}.  
\begin{rem}
While the indexing used in \cite{pereira2022tensor} for the moments  is slightly different from ours,
it is easy to convert between the two. 
Let  $m_{v_1\cdots v_d}$ be a $t$-th order multivariate moment and let $M_{v_1\cdots v_d}^{(t)}$ be an entry of the $t$-th order tensor moment.  Then $m_{v_1\cdots v_d}=M_{a_1\cdots a_d}^{(t)}$ where 
$a_j=|\{v_k=j\}|$ and $t=\sum_{j=1}^n v_j$.  Note that due to symmetry, the indexing of the tensor moment is non-unique.  For example, $m_{102} = M_{133}^{(3)}=M_{331}^{(3)}=M_{313}^{(3)}$. 
\hfill$\diamond$\end{rem}

The closed form tensor moment formula has some drawbacks in implementation, which is why \cite{pereira2022tensor} bypassed generating the moment tensors in their optimization-based estimation method.  Moment tensors are large, and the symmetry of tensor moments creates a great deal of duplication, which then uses a great deal of memory, limiting the size of problems that can be addressed.  The symmetry also results from symmetrizing a non-symmetric tensor which requires summing over the permutations of the indices of each entry.  This can be somewhat mitigated by collecting only one copy of this sum and only for the entries we need.  However as $k$ values grow, the necessary permutation becomes unmanageable.  This suggests that more computationally efficient methods of generating the mixed dimensional moment polynomials would be of great help in solving these problems.  Nevertheless this formulation allows us to generate the linear system that solves for the off-diagonal covariance entries. 

\subsubsection{Recursive Method}
Alternatively, as noted in \cite{lindberg2023estimating} we can reduce the multivariate moments we are concerned with to single variate moments via the equivalence $$m_{te_i+e_j}(\lambda_t,\mu_t,\sigma_t)=\sum_{\ell=1}^k\lambda_\ell(\mu_{\ell j}\cdot m_t(\mu_{\ell i},\sigma_{\ell i i})+t\sigma_{\ell i j}\cdot m_{t-1}(\mu_{\ell i}, \sigma_{\ell i i}))$$ where $m_t(\mu_{\ell i},\sigma_{\ell i i})$ denotes the $t$-th moment of a Gaussian random variable with mean $\mu_{\ell i}$ and variance $\sigma_{\ell i i}$.  We can then use the usual recursive method for generating the relevant moment polynomials.

\begin{rem}
    For parameter recovery with known $\lambda$, we can use moments $$V_2:=\{\overline{m}_{e_i},\overline{m}_{2e_i},\dots,\overline{m}_{(2k+1)e_i}\}$$ for $i\in[d]$ and $V_3$ as described above.  Then carry out Steps 3 - 6 of Algorithm \ref{alg:general} where Step 3 is additionally completed for $i=1$.
\end{rem}

\section{Computational Results and Benchmarks}\label{sec:computational}

All timings were completed using an Intel Xeon e5-2698 with 250GiB of RAM.  Implementation code is available via the \texttt{julia} package GMMParameterEstimation and  
\begin{center}
    \url{github.com/HaleyColgateKottler/GMMParameterEstimation.jl} 
\end{center}
where the polynomial systems are solved using \\
\texttt{HomotopyContinuation.jl} \cite{homotopyContinuation}.  Tables of estimates used to generate the figures can be found in Appendix \ref{app:ests}.

\begin{figure}
    \centering
    \includegraphics[width=\linewidth]{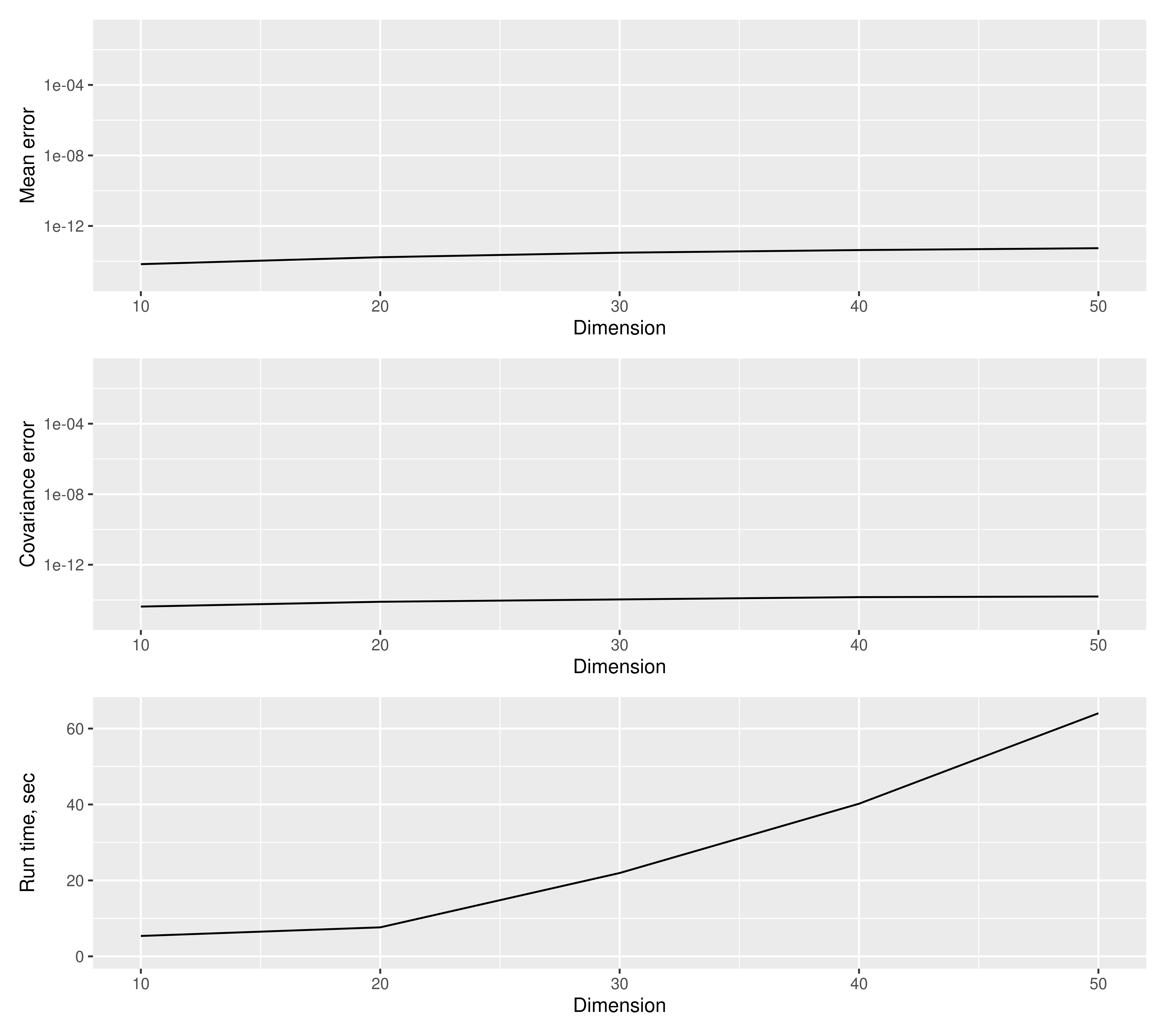}
    \Description{Graphs showing increasing errors as dimension increases}
    \caption{(Exact moments, mixing coefficients known)
    Median mean, and covariance errors for exact moments over 1000 random test runs with known mixing coefficients normalized by the number of parameters.}
    \label{fig:perfect-moments-general-cov-known-mixing}
\end{figure}

\begin{figure}
    \centering
    \includegraphics[width=\linewidth]{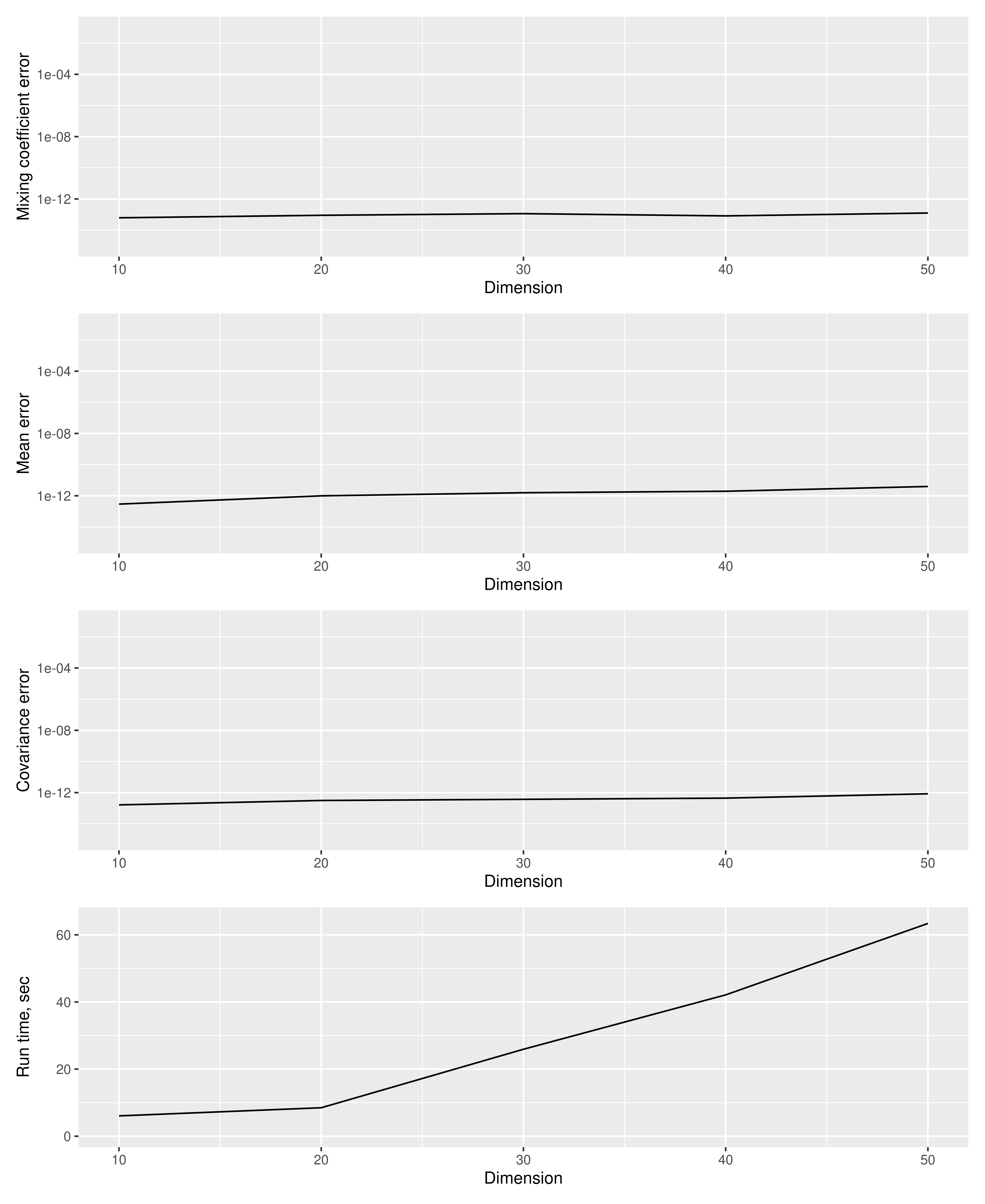}
    \Description{Graphs showing increasing errors and run times as dimension increases}
    \caption{(Exact moments, $\Sigma_\ell\succ0$)
    Median mixing coefficient, mean, and covariance errors for exact moments over 1000 random test runs with unknown mixing coefficients normalized by the number of parameters.}
    \label{fig:perfect-moments-general-cov-unknown-mixing}
\end{figure}

\subsection{Methods}
In order to establish benchmarks for future works and enable replication of our results, we provide the details of our test framework.
\subsubsection{Generating parameters}
To apply our implementation,
we generated random Gaussian mixture models via synthetic parameters. 
We generated mixing coefficients, means, and covariances as follows.  To generate the mixing coefficients, we created a vector of $k$ entries by selecting each entry independently from a $\mathcal{N}(0,1)$ distribution, took the element-wise absolute value of this vector, then scaled scaled the vector by its sum.  To generate the means we took $k\times d$ samples from $\mathcal{N}(0,1)$ and arranged them in a matrix to represent each mean vector. Finally, to generate each covariance matrix $\Sigma_j$, $j\in[k]$, we generated a matrix $M\in\mathbb{R}^{d\times d}$ by sampling each entry independently from a $\mathcal{N}(0,1)$ distribution, and took $ \Sigma_j = M\cdot M^T$.  If diagonal covariance matrices were desired, we then took just the diagonal of each covariance matrix.

\subsubsection{Error computation}
Recall that there are label-swapping symmetries inherent to 
the problem of estimating the parameters of mixture models. 
In each experiment we produced below, errors are computed by 
ordering components to minimize the Euclidean norm between the true and estimated mixing coefficients. 
We then take the Euclidean norm between each component of the true and estimated parameters, divided by the number of entries to get a normalized error.  See Appendix \ref{app:error} for code to allow for replication.

\subsection{Exact Moments}
As a baseline, we first tested our method using exact moments to demonstrate that because the Method of Moments is consistent with accurate moments we achieve machine precision.  By evaluating the moment functions $f_v(\theta)$ at the exact parameters, 
we arrive at machine accuracy exact moments.  We then use our implementation to recover parameters, which allows for analysis of the error introduced via the algorithm. As seen in Tables \ref{fig:perfect-moments-general-cov-unknown-mixing} and \ref{fig:perfect-moments-general-cov-known-mixing}, we achieve  machine precision for both the known and unknown mixing coefficient cases with three components and non-diagonal covariance matrices.  This supports the argument that the majority of the error introduced is from computing the moments from samples, not from the algorithm.

\begin{figure}
    \centering
    \includegraphics[width=\linewidth]{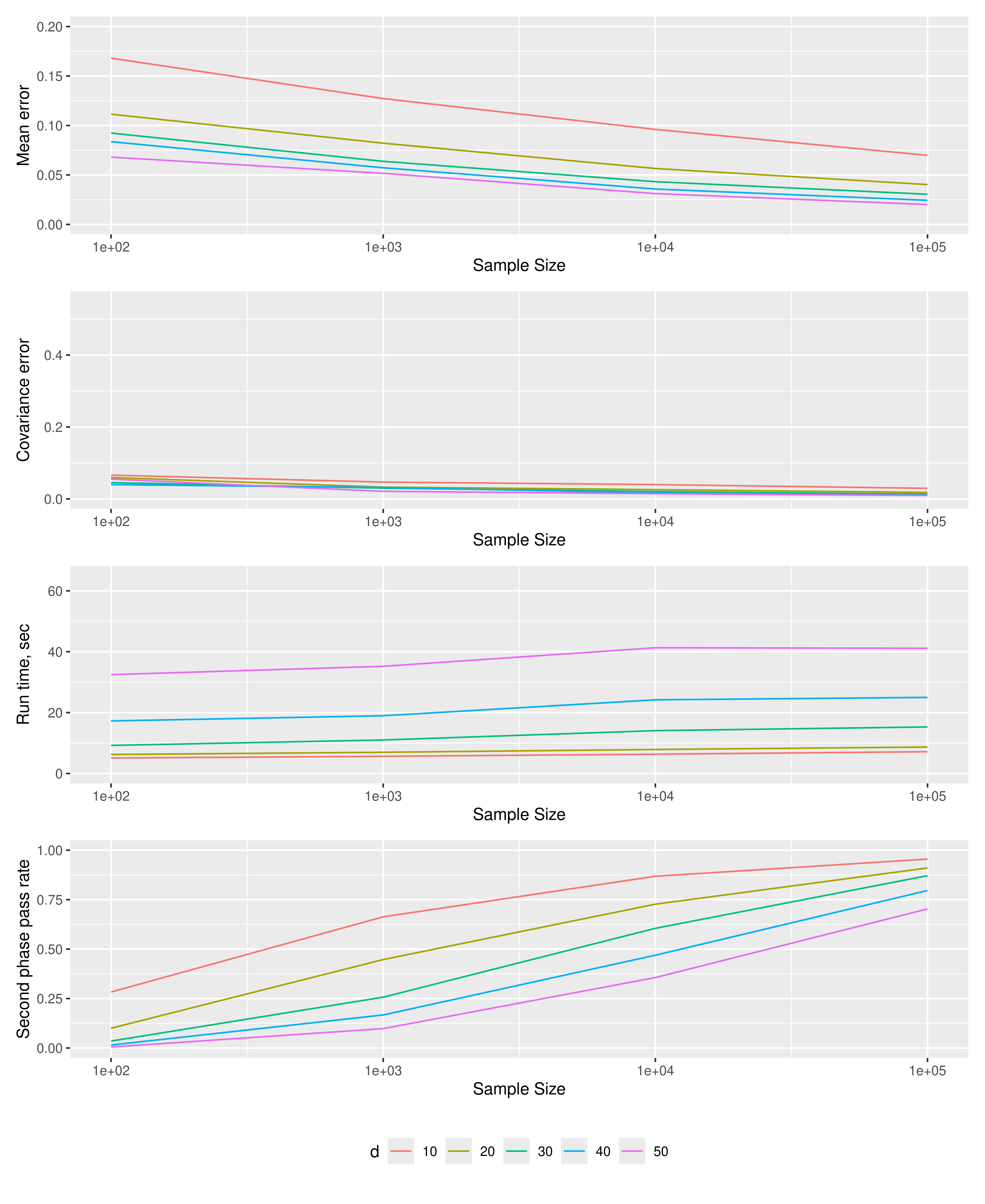}
    \Description{Graphs showing decreasing errors and increasing pass rates and run times as sample size increases}
    \caption{(Sample moments, mixing coefficients known and $\Sigma_\ell\succ0$) Median mean, and covariance errors, and number of statistically meaningful solutions for sample moments over 1000 random test runs with known mixing coefficients with $k = 3$ normalized by the number of parameters.}\label{fig:sample-moments-general-cov-known-mixing}
\end{figure}

\subsection{Sample Moments}
We also test our implementation using samples generated from the true distribution.  Our implementation is flexible in using both the lower and higher order recursive systems, as demonstrated in Figure \ref{fig:kvslowsystems}.  We expect that for increased $k$, the distinction between the methods will become clearer, but for consistency we used the lower order system for our tests. As seen in Tables \ref{fig:perfect-moments-general-cov-unknown-mixing} and \ref{fig:sample-moments-general-cov-known-mixing}, we see that error decreases as expected with more samples or fewer dimensions overall.  We also see that the majority of cases where the algorithm is unable to find a statistically meaningful solution, the issue is in solving the first system for mixing coefficients and the means and covariances for the first dimension.  In sample data, if another dimension leads to more accurate moments, using that dimension to recover mixing coefficients and then proceeding can address this difficulty.  We therefore propose a modification to Algorithm 1 as Algorithm 2.  This cycling can greatly increase the likelihood of finding a statistically meaningful solution.

\subsection{Benchmark: Uniform Mixtures with Known Shared Covariance Matrices}
\begin{algorithm}[htb]
\SetKwInOut{Input}{Input}\SetKwInOut{Output}{Output}
\caption{Density Estimation for Uniform Mixtures of Multivariate Gaussians with Equal Covariances \cite{lindberg2023estimating}}\label{alg:sharedKnown}
\Input{Sample covariance $\Sigma$, and the set of sample moments: $$V_1:=\{\overline{m}_{e_1},\dots,\overline{m}_{ke_1}\}$$ $$V_2:=\{\overline{m}_{e_i},\overline{m}_{e_1+e_i},\overline{m}_{2e_1+e_i},\dots,\overline{m}_{(k-1)e_1+e_i}\}_{i=2}^d$$ that are the moments to multivariate Gaussian mixture model: $$\frac{1}{k}\mathcal{N}(\mu_1,\Sigma)+\cdots+\frac{1}{k}\mathcal{N}(\mu_k,\Sigma).$$
}
\Output{ Parameters $\mu_\ell\in\mathbb{R}^d$, such that $V_1\cup V_2$, are the moments of distribution $\sum_{\ell=1}^k\frac{1}{k}\mathcal{N}(\mu_\ell,\Sigma)$.
}
\BlankLine
Using mixing coefficients $\lambda_\ell=\frac{1}{k}$ for $\ell\in[k]$ and sample moments $V_1$ solve the univariate system to obtain $\mu_{\ell 1}\in\mathbb{R}$.\; 
Using sample moments $V_2$ solve the $k\times k$ linear system in $\mu_{i1},\cdots,\mu_{ik}$ for $2\leq i\leq d$.
\end{algorithm}

\begin{figure}
    \centering
    \includegraphics[width=\linewidth]{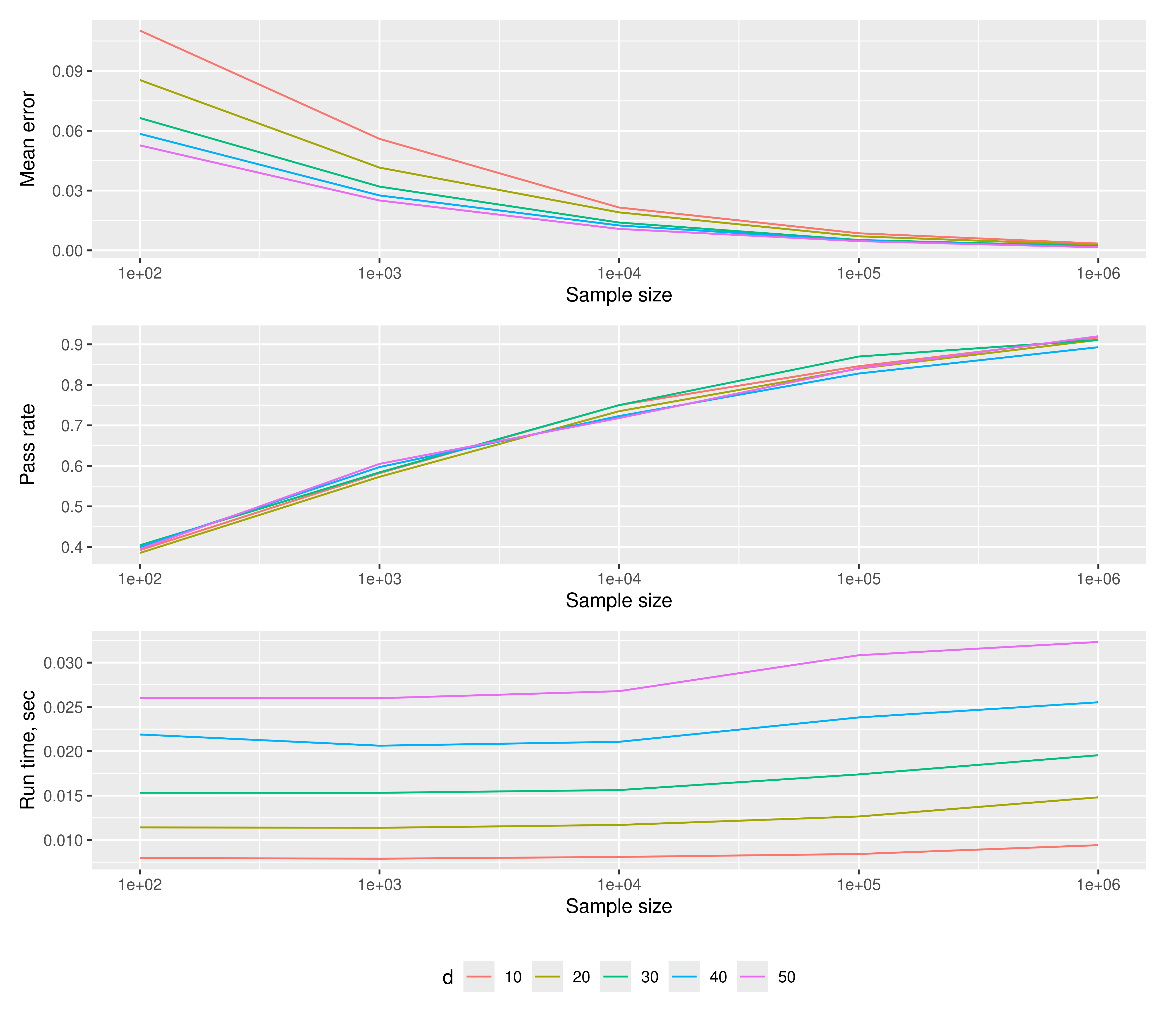}
    \Description{Graphs showing decreasing errors and increasing pass rates and run times as sample size increases}
    \caption{(Sample moments, $\lambda_\ell=\frac{1}{k}$ and $\Sigma_\ell=I$)
    Median mean errors, and number of statistically meaningful solutions for sample moments over 1000 random test runs with identity covariances and equal mixing coefficients with $k = 3$ normalized by the number of parameters.}
    \label{fig:equal-cov-known-equal-mixing}
\end{figure}

As previously discussed, the parameter recovery problem \ref{prob:param_recovery} for Gaussian mixture models is well studied.  Recently, most work focuses on the case where each component has the same known covariance matrix and the mixture is uniform \cite{daskalakis2017ten,xu2016global}.  Our implementation includes this case as well to match the literature. 
For this final benchmark,
we implemented Algorithm \ref{alg:sharedKnown} from \cite{lindberg2023estimating}. As seen in Figure \ref{fig:equal-cov-known-equal-mixing}, we obtain significantly higher rates of finding a statistically meaningful solution with the added restriction of uniform mixtures and equal covariances.  Overall error is low even with small sample sizes.

\section{Acknowledgments}
Research of Haley Colgate Kottler was supported in part by NSF Award DMS-2023239.
Research of Jose Israel Rodriguez is supported by the Sloan Foundation and the Office of the Vice Chancellor for Research and Graduate Education at U.W. Madison with funding from the Wisconsin Alumni Research Foundation.

\bibliographystyle{plain}
\bibliography{ref}

\appendix 
\section{Error Computation}\label{app:error}
\begin{leftbar}
\begin{verbatim}
using Combinatorics
function computeError(w, true_means, true_covars,
                      mixcoefs, means, covars,
                      diagonal)
    k, d = size(true_means)

    weight_errs = [norm(mixcoefs[
                            nthperm(1:k, i),
                            :] - w) 
                            for i in 1:factorial(k)]
    min_weight_err, best_perm = findmin(weight_errs)

    final_mixcoefs = mixcoefs[
                                nthperm(1:k,
                                                    best_perm), :]
    final_means = means[nthperm(1:k, best_perm), :]

    if diagonal
        final_covars = [covars[i][1:end, 1:end] 
                             for i in nthperm(1:k, best_perm)]
    else
        final_covars = covars[nthperm(1:k,
                                        best_perm),
                                        :, :]
    end

    return (norm(final_mixcoefs - w), 
            norm(final_means - true_means), 
            norm(final_covars - true_covars))
end
\end{verbatim}
\end{leftbar}

\begin{table}[h]
\begin{tabular}{|cccc|}\hline
\multicolumn{1}{|c|}{d}     & \multicolumn{1}{c|}{Mean Error} & \multicolumn{1}{c|}{Covariance Error} & \begin{tabular}[c]{@{}l@{}}Time in Seconds\end{tabular} \\ \hline
\multicolumn{1}{|c|}{10}    & \multicolumn{1}{c|}{$6.81\times10^{-15}$}      & \multicolumn{1}{c|}{$4.22\times10^{-15}$}            & 5.37                                                                         \\ \cline{1-1}
\multicolumn{1}{|c|}{20}    & \multicolumn{1}{c|}{$1.68\times10^{-14}$}      & \multicolumn{1}{c|}{$7.91\times10^{-15}$}            & 7.64                                                                       \\ \cline{1-1}
\multicolumn{1}{|c|}{30}    & \multicolumn{1}{c|}{$3.03\times10^{-14}$}      & \multicolumn{1}{c|}{$1.07\times10^{-14}$}            & 21.95                                                           \\ \cline{1-1}
\multicolumn{1}{|c|}{40}    & \multicolumn{1}{c|}{$4.28\times10^{-14}$}      & \multicolumn{1}{c|}{$1.45\times10^{-14}$}            & 40.20
\\ \cline{1-1}
\multicolumn{1}{|c|}{50}    & \multicolumn{1}{c|}{$5.46\times10^{-14}$}      & \multicolumn{1}{c|}{$1.57\times10^{-14}$}            & 64.03   \\ \hline
\end{tabular}
\caption{
(Exact moments, mixing coefficients known)
Median mean, and covariance errors for exact moments over 1000 random test runs with known mixing coefficients normalized by the number of parameters.}
\label{table:perfect-moments-general-cov-known-mixing}
\end{table}
\newpage 

\section{Numerical Results}\label{app:ests}
See Tables \ref{table:perfect-moments-general-cov-known-mixing} and \ref{table:perfect-moments-general-cov-unknown-mixing} for results with perfect moments and Tables \ref{table:sample-moments-general-cov-known-mixing}, \ref{table:sample-moments-general-cov-unknown-mixing}, and \ref{table:equal-cov-known-equal-mixing} for results from sample moments.

\begin{table}[hb]
\begin{tabular}{|ccccc|}\hline
\multicolumn{1}{|c|}{d}      & \multicolumn{1}{c|}{Mixing Coefficient} & \multicolumn{1}{c|}{Mean Error} & \multicolumn{1}{c|}{Covariance Error} & \begin{tabular}[c]{@{}l@{}}Time (s)\end{tabular} \\ \hline
\multicolumn{1}{|c|}{10}    & \multicolumn{1}{c|}{$6.17\times10^{-14}$} & \multicolumn{1}{c|}{$2.93\times10^{-13}$} & \multicolumn{1}{c|}{$1.63\times10^{-13}$} & 6.06                                                                             \\ \cline{1-1}
\multicolumn{1}{|c|}{20}    & \multicolumn{1}{c|}{$8.95\times10^{-14}$} & \multicolumn{1}{c|}{$9.93\times10^{-13}$} & \multicolumn{1}{c|}{$3.13\times10^{-13}$} & 8.48                                                                        \\ \cline{1-1}
\multicolumn{1}{|c|}{30}    & \multicolumn{1}{c|}{$1.14\times10^{-13}$} & \multicolumn{1}{c|}{$1.57\times10^{-12}$} & \multicolumn{1}{c|}{$3.71\times10^{-13}$} & 25.91                                                                          \\ \cline{1-1}
\multicolumn{1}{|c|}{40}    & \multicolumn{1}{c|}{$8.19\times10^{-14}$} & \multicolumn{1}{c|}{$1.96\times10^{-12}$} & \multicolumn{1}{c|}{$4.44\times10^{-13}$} & 42.11     
\\ \cline{1-1}
\multicolumn{1}{|c|}{50}    & \multicolumn{1}{c|}{$1.25\times10^{-13}$} & \multicolumn{1}{c|}{$3.96\times10^{-12}$} & \multicolumn{1}{c|}{$8.43\times10^{-13}$} & 63.39                                                                        \\ \hline
\end{tabular}
\caption{(Exact moments, $\Sigma_\ell\succ0$)
Median mixing coefficient, mean, and covariance errors for exact moments over 1000 random test runs with unknown mixing coefficients normalized by the number of parameters.}\label{table:perfect-moments-general-cov-unknown-mixing}
\end{table}

\begin{table}[hb]
\begin{tabular}{|clllcc|}
\hline
\multicolumn{6}{|l|}{Median Mean Error}                                                                                  \\ \hline
\multicolumn{1}{|l|}{d}               & \multicolumn{1}{c|}{10} & \multicolumn{1}{c|}{20} & \multicolumn{1}{c|}{30} & \multicolumn{1}{c|}{40} & 50\\ \hline
\multicolumn{1}{|l|}{100 samples}     & \multicolumn{1}{c|}{0.168}   & \multicolumn{1}{c|}{0.111}   & \multicolumn{1}{c|}{0.092}   & \multicolumn{1}{c|}{0.084} & 0.068  \\ \cline{1-1}
\multicolumn{1}{|l|}{1000 samples}    & \multicolumn{1}{c|}{0.127}   & \multicolumn{1}{c|}{0.082}   & \multicolumn{1}{c|}{0.064}   & \multicolumn{1}{c|}{0.057} & 0.052  \\ \cline{1-1}
\multicolumn{1}{|l|}{10,000 samples}  & \multicolumn{1}{c|}{0.096}   & \multicolumn{1}{c|}{0.057}   & \multicolumn{1}{c|}{0.043}   & \multicolumn{1}{c|}{0.036} & 0.031  \\ \cline{1-1}
\multicolumn{1}{|l|}{100,000 samples} & \multicolumn{1}{c|}{0.070}   & \multicolumn{1}{c|}{0.040}   & \multicolumn{1}{c|}{0.030}   & \multicolumn{1}{c|}{0.024} & 0.020  \\ \hline
\multicolumn{6}{|l|}{Median Covariance Error}                                                                                  \\ \hline
\multicolumn{1}{|l|}{d}               & \multicolumn{1}{c|}{10} & \multicolumn{1}{c|}{20} & \multicolumn{1}{c|}{30} & \multicolumn{1}{c|}{40} & 50\\ \hline
\multicolumn{1}{|l|}{100 samples}     & \multicolumn{1}{c|}{0.0662}   & \multicolumn{1}{c|}{0.0595}   & \multicolumn{1}{c|}{0.0454}   & \multicolumn{1}{c|}{0.0397} & 0.0550  \\ \cline{1-1}
\multicolumn{1}{|l|}{1000 samples}    & \multicolumn{1}{c|}{0.0469}   & \multicolumn{1}{c|}{0.0330}   & \multicolumn{1}{c|}{0.0300}   & \multicolumn{1}{c|}{0.0319} & 0.0212  \\ \cline{1-1}
\multicolumn{1}{|l|}{10,000 samples}  & \multicolumn{1}{c|}{0.0399}   & \multicolumn{1}{c|}{0.0258}   & \multicolumn{1}{c|}{0.0205}   & \multicolumn{1}{c|}{0.0167} & 0.0145  \\ \cline{1-1}
\multicolumn{1}{|l|}{100,000 samples} & \multicolumn{1}{c|}{0.0294}   & \multicolumn{1}{c|}{0.0182}   & \multicolumn{1}{c|}{0.0133}   & \multicolumn{1}{c|}{0.0105} & 0.0096  \\ \hline
\multicolumn{6}{|l|}{Pass Rate}                                                                                  \\ \hline
\multicolumn{1}{|l|}{d}               & \multicolumn{1}{c|}{10} & \multicolumn{1}{c|}{20} & \multicolumn{1}{c|}{30} & \multicolumn{1}{c|}{40} & 50\\ \hline
\multicolumn{1}{|l|}{100 samples}     & \multicolumn{1}{c|}{283}   & \multicolumn{1}{c|}{100}   & \multicolumn{1}{c|}{35}   & \multicolumn{1}{c|}{15} & 5  \\ \cline{1-1}
\multicolumn{1}{|l|}{1000 samples}    & \multicolumn{1}{c|}{663}   & \multicolumn{1}{c|}{447}   & \multicolumn{1}{c|}{257}   & \multicolumn{1}{c|}{167} & 98  \\ \cline{1-1}
\multicolumn{1}{|l|}{10,000 samples}  & \multicolumn{1}{c|}{868}   & \multicolumn{1}{c|}{727}   & \multicolumn{1}{c|}{605}   & \multicolumn{1}{c|}{469} & 356  \\ \cline{1-1}
\multicolumn{1}{|l|}{100,000 samples} & \multicolumn{1}{c|}{955}   & \multicolumn{1}{c|}{910}   & \multicolumn{1}{c|}{871}   & \multicolumn{1}{c|}{796} & 703  \\ \hline
\end{tabular}
\caption{(Sample moments, mixing coefficients known and $\Sigma_\ell\succ0$) Median mean, and covariance errors, and number of statistically meaningful solutions for sample moments over 1000 random test runs with known mixing coefficients with $k = 3$ normalized by the number of parameters.}\label{table:sample-moments-general-cov-known-mixing}
\end{table}

\newpage 
\begin{table}[hb]
\begin{tabular}{|cccccc|}
\hline
\multicolumn{6}{|l|}{Median Mixing Coefficient Error}  \\ \hline
\multicolumn{1}{|l|}{d} & \multicolumn{1}{c|}{10} & \multicolumn{1}{c|}{20} & \multicolumn{1}{c|}{30} & \multicolumn{1}{|l|}{40} & 50 \\ \hline
\multicolumn{1}{|l|}{100 samples}     & \multicolumn{1}{c|}{0.0507}  & \multicolumn{1}{c|}{0.0338}   & \multicolumn{1}{c|}{0.0208}   & \multicolumn{1}{c|}{0.0216} & 0.0229\\ \cline{1-1}
\multicolumn{1}{|l|}{1000 samples}    & \multicolumn{1}{c|}{0.0267}  & \multicolumn{1}{c|}{0.0187}   & \multicolumn{1}{c|}{0.0114}   & \multicolumn{1}{c|}{0.0103} & 0.0056  \\ \cline{1-1}
\multicolumn{1}{|l|}{10,000 samples}  & \multicolumn{1}{c|}{0.0249}  & \multicolumn{1}{c|}{0.0069}   & \multicolumn{1}{c|}{0.0046}   & \multicolumn{1}{c|}{0.0041} & 0.0030  \\ \cline{1-1}
\multicolumn{1}{|l|}{100,000 samples} & \multicolumn{1}{c|}{0.0160}  & \multicolumn{1}{c|}{0.0058}   & \multicolumn{1}{c|}{0.0025}   & \multicolumn{1}{c|}{0.0016} & 0.0015  \\ \hline
\multicolumn{6}{|l|}{Median Mean Error}                                                                                  \\ \hline
\multicolumn{1}{|l|}{d}               & \multicolumn{1}{c|}{10}      & \multicolumn{1}{c|}{20}      & \multicolumn{1}{c|}{30} & \multicolumn{1}{c|}{40} & 50 \\ \hline
\multicolumn{1}{|l|}{100 samples}     & \multicolumn{1}{c|}{0.186}   & \multicolumn{1}{c|}{0.143}   & \multicolumn{1}{c|}{0.091}   & \multicolumn{1}{c|}{0.068} & 0.092  \\ \cline{1-1}
\multicolumn{1}{|l|}{1000 samples}    & \multicolumn{1}{c|}{0.145}   & \multicolumn{1}{c|}{0.093}   & \multicolumn{1}{c|}{0.066}   & \multicolumn{1}{c|}{0.053} & 0.064  \\ \cline{1-1}
\multicolumn{1}{|l|}{10,000 samples}  & \multicolumn{1}{c|}{0.133}   & \multicolumn{1}{c|}{0.077}   & \multicolumn{1}{c|}{0.060}   & \multicolumn{1}{c|}{0.041} & 0.037  \\ \cline{1-1}
\multicolumn{1}{|l|}{100,000 samples} & \multicolumn{1}{c|}{0.099}   & \multicolumn{1}{c|}{0.054}   & \multicolumn{1}{c|}{0.033}   & \multicolumn{1}{c|}{0.028} & 0.022  \\ \hline
\multicolumn{6}{|l|}{Median Covariance Error}                                                                                  \\ \hline
\multicolumn{1}{|l|}{d}               & \multicolumn{1}{c|}{10}      & \multicolumn{1}{c|}{20}      & \multicolumn{1}{c|}{30} & \multicolumn{1}{c|}{40} & 50 \\ \hline
\multicolumn{1}{|l|}{100 samples}     & \multicolumn{1}{c|}{0.526}   & \multicolumn{1}{c|}{0.023}   & \multicolumn{1}{c|}{0.079}   & \multicolumn{1}{c|}{0.088} & 0.014  \\ \cline{1-1}
\multicolumn{1}{|l|}{1000 samples}    & \multicolumn{1}{c|}{0.056}   & \multicolumn{1}{c|}{0.041}   & \multicolumn{1}{c|}{0.021}   & \multicolumn{1}{c|}{0.037} & 0.026  \\ \cline{1-1}
\multicolumn{1}{|l|}{10,000 samples}  & \multicolumn{1}{c|}{0.053}   & \multicolumn{1}{c|}{0.026}   & \multicolumn{1}{c|}{0.022}   & \multicolumn{1}{c|}{0.019} & 0.019  \\ \cline{1-1}
\multicolumn{1}{|l|}{100,000 samples} & \multicolumn{1}{c|}{0.034}   & \multicolumn{1}{c|}{0.021}   & \multicolumn{1}{c|}{0.014}   & \multicolumn{1}{c|}{0.012} & 0.009  \\ \hline
\multicolumn{6}{|l|}{Pass Rate}                                                                                  \\ \hline
\multicolumn{1}{|l|}{d}               & \multicolumn{1}{c|}{10} & \multicolumn{1}{c|}{20} & \multicolumn{1}{c|}{30} & \multicolumn{1}{c|}{40} & 50 \\ \hline
\multicolumn{1}{|l|}{100 samples}     & \multicolumn{1}{c|}{4}   & \multicolumn{1}{c|}{5}   & \multicolumn{1}{c|}{4}   & \multicolumn{1}{c|}{2} & 1  \\ \cline{1-1}
\multicolumn{1}{|l|}{1000 samples}    & \multicolumn{1}{c|}{58}   & \multicolumn{1}{c|}{39}   & \multicolumn{1}{c|}{23}   & \multicolumn{1}{c|}{17} & 7  \\ \cline{1-1}
\multicolumn{1}{|l|}{10,000 samples}  & \multicolumn{1}{c|}{140}   & \multicolumn{1}{c|}{79}   & \multicolumn{1}{c|}{59}   & \multicolumn{1}{c|}{46} & 30  \\ \cline{1-1}
\multicolumn{1}{|l|}{100,000 samples} & \multicolumn{1}{c|}{219}   & \multicolumn{1}{c|}{162}   & \multicolumn{1}{c|}{99}   & \multicolumn{1}{c|}{94} & 77  \\ \hline
\multicolumn{6}{|l|}{First Dimension Failure}                                                                                  \\ \hline
\multicolumn{1}{|l|}{d}               & \multicolumn{1}{c|}{10} & \multicolumn{1}{c|}{20} & \multicolumn{1}{c|}{30} & \multicolumn{1}{c|}{40} & 50\\ \hline
\multicolumn{1}{|l|}{100 samples}     & \multicolumn{1}{c|}{918}   & \multicolumn{1}{c|}{882}   & \multicolumn{1}{c|}{885}   & \multicolumn{1}{c|}{876} & 885  \\ \cline{1-1}
\multicolumn{1}{|l|}{1000 samples}    & \multicolumn{1}{c|}{741}   & \multicolumn{1}{c|}{739}   & \multicolumn{1}{c|}{743}   & \multicolumn{1}{c|}{752} & 716  \\ \cline{1-1}
\multicolumn{1}{|l|}{10,000 samples}  & \multicolumn{1}{c|}{525}   & \multicolumn{1}{c|}{576}   & \multicolumn{1}{c|}{559}   & \multicolumn{1}{c|}{541} & 577  \\ \cline{1-1}
\multicolumn{1}{|l|}{100,000 samples} & \multicolumn{1}{c|}{400}   & \multicolumn{1}{c|}{401}   & \multicolumn{1}{c|}{396}   & \multicolumn{1}{c|}{413} & 407  \\ \hline
\end{tabular}
\caption{(Sample moments, $\Sigma_\ell\succ0$)
Median mixing coefficient, mean, and covariance errors, and number of statistically meaningful solutions for sample moments over 1000 random test runs with unknown mixing coefficients with $k = 3$ normalized by the number of parameters.}\label{table:sample-moments-general-cov-unknown-mixing}
\end{table}

\newpage
\begin{table}[hb]
\begin{tabular}{|clllcc|}
\hline
\multicolumn{6}{|l|}{Median Mean Error}                                                                                  \\ \hline
\multicolumn{1}{|l|}{d}                 & \multicolumn{1}{c|}{10} & \multicolumn{1}{c|}{20} & \multicolumn{1}{c|}{30} & \multicolumn{1}{c|}{40} & 50\\ \hline
\multicolumn{1}{|l|}{100 samples}       & \multicolumn{1}{c|}{0.110}   & \multicolumn{1}{c|}{0.085}   & \multicolumn{1}{c|}{0.066}   & \multicolumn{1}{c|}{0.058} & 0.053  \\ \cline{1-1}
\multicolumn{1}{|l|}{1000 samples}      & \multicolumn{1}{c|}{0.056}   & \multicolumn{1}{c|}{0.042}   & \multicolumn{1}{c|}{0.032}   & \multicolumn{1}{c|}{0.028} & 0.025  \\ \cline{1-1}
\multicolumn{1}{|l|}{10,000 samples}    & \multicolumn{1}{c|}{0.021}   & \multicolumn{1}{c|}{0.019}   & \multicolumn{1}{c|}{0.014}   & \multicolumn{1}{c|}{0.013} & 0.011  \\ \cline{1-1}
\multicolumn{1}{|l|}{100,000 samples}   & \multicolumn{1}{c|}{0.009}   & \multicolumn{1}{c|}{0.007}   & \multicolumn{1}{c|}{0.005}   & \multicolumn{1}{c|}{0.005} & 0.005  \\ \cline{1-1}
\multicolumn{1}{|l|}{1,000,000 samples} & \multicolumn{1}{c|}{0.003}   & \multicolumn{1}{c|}{0.003}   & \multicolumn{1}{c|}{0.002}   & \multicolumn{1}{c|}{0.002} & 0.002  \\ \hline
\multicolumn{6}{|l|}{Pass Rate}                                                                                  \\ \hline
\multicolumn{1}{|l|}{d}                 & \multicolumn{1}{c|}{10} & \multicolumn{1}{c|}{20} & \multicolumn{1}{c|}{30} & \multicolumn{1}{c|}{40} & 50\\ \hline
\multicolumn{1}{|l|}{100 samples}       & \multicolumn{1}{c|}{392}   & \multicolumn{1}{c|}{385}   & \multicolumn{1}{c|}{404}   & \multicolumn{1}{c|}{400} & 395  \\ \cline{1-1}
\multicolumn{1}{|l|}{1000 samples}      & \multicolumn{1}{c|}{582}   & \multicolumn{1}{c|}{573}   & \multicolumn{1}{c|}{584}   & \multicolumn{1}{c|}{597} & 605  \\ \cline{1-1}
\multicolumn{1}{|l|}{10,000 samples}    & \multicolumn{1}{c|}{750}   & \multicolumn{1}{c|}{735}   & \multicolumn{1}{c|}{750}   & \multicolumn{1}{c|}{723} & 718  \\ \cline{1-1}
\multicolumn{1}{|l|}{100,000 samples}   & \multicolumn{1}{c|}{846}   & \multicolumn{1}{c|}{840}   & \multicolumn{1}{c|}{870}   & \multicolumn{1}{c|}{828} & 841  \\ \cline{1-1}
\multicolumn{1}{|l|}{1,000,000 samples} & \multicolumn{1}{c|}{917}   & \multicolumn{1}{c|}{911}   & \multicolumn{1}{c|}{911}   & \multicolumn{1}{c|}{893} & 920  \\ \hline
\end{tabular}
\caption{(Sample moments, $\lambda_\ell=\frac{1}{k}$ and $\Sigma_\ell=I$)
Median mean errors, and number of statistically meaningful solutions for sample moments over 1000 random test runs with identity covariances and equal mixing coefficients with $k = 3$ normalized by the number of parameters.}\label{table:equal-cov-known-equal-mixing}
\end{table}
\end{document}